\documentclass[letterpaper,11pt,fleqn]{article}
\usepackage{jheppub}
\usepackage{bm,amsmath,amssymb}

\long\def\comment#1{ }

\newcommand{\eqn}[1]{Eq.~\eqref{#1}}

\newcommand{\beq}{\begin{eqnarray}}
\newcommand{\eeq}{\end{eqnarray}}
\newcommand{\nn}{\nonumber\\}

\newcommand{\dif}{{\rm d}}
\newcommand{\rmd}{{\rm d}}
\newcommand{\rme}{{\rm e}}
\newcommand{\rmi}{{\rm i}}

\newcommand{\rmtr}{{\rm tr}}

\newcommand{\del}{\partial}

\newcommand{\lan}{\langle}
\newcommand{\ran}{\rangle}

\newcommand{\mcal}{\mathcal}
\newcommand{\wt}{\widetilde}

\newcommand{\bmx}{\bm{x}}
\newcommand{\bmy}{\bm{y}}
\newcommand{\bmu}{\bm{u}}
\newcommand{\bmv}{\bm{v}}
\newcommand{\bmz}{\bm{z}}

\newcommand{\abar}{\bar{\alpha}}
\newcommand{\atpi}{\frac{\abar}{2 \pi}}

\let\Oldcdot\cdot
\renewcommand{\cdot}{\mspace{-2mu}\Oldcdot\mspace{-2mu}}
\let\Oldtimes\times
\renewcommand{\times}{\mspace{-2mu}\Oldtimes\mspace{-2mu}}

\title{\Large Higher--point correlations from the JIMWLK evolution}

\author[a]{E.~Iancu}
\author[b]{and D.N.~Triantafyllopoulos}

\affiliation[a]{Institut de Physique Th\'{e}orique de Saclay,
F-91191 Gif-sur-Yvette, France}
\affiliation[b]{ECT*, European Center for Theoretical Studies in Nuclear
Physics and Related Areas,\\ Strada delle Tabarelle 286, I-38123 Villazzano
(TN), Italy}

\emailAdd{edmond.iancu@cea.fr}
\emailAdd{trianta@ectstar.eu}

\abstract{We develop a new approximation scheme aiming at extracting higher--point correlation functions from the JIMWLK evolution, in the limit where the
number of colors is large. Namely, we show that by exploiting the
structure of the `virtual' terms in the Balitsky--JIMWLK equations, one
can derive functional relations expressing arbitrary $n$--point functions
of the Wilson lines in terms of the 2--point function (the scattering
amplitude for a color dipole). These approximations are correct
not only in the regime of strong scattering, where the evolution is
indeed controlled by the `virtual' terms, but also in the regime of weak
scattering, where they reduce to the corresponding BFKL solutions.
This last feature follows from the fact that the JIMWLK Hamiltonian
is a linear combination of the pieces responsible for the `real' and
`virtual' terms, respectively. We apply this scheme to two examples:
the `color quadrupole' (the 4--point function of the Wilson lines
which enters the cross--section for the production of a pair of jets at forward rapidities) and the `color sextupole' (the 6--point function). For particular
configurations of the quadrupole, our general formula reduces
to relatively simple expressions that have been previously proposed on
the basis of the McLerran--Venugopalan model and which were recently
shown to agree quite well with exact, numerical, solutions to the JIMWLK
equation.
}
\arxivnumber{1109.0302}
\vfill
\begin{document}
\maketitle

\section{Introduction}
\label{sect:intro}

Since the derivation, more than a decade ago, of the equations describing
the high--energy evolution in QCD to leading logarithmic accuracy --- the
infinite hierarchy of Balitsky equations \cite{Balitsky:1995ub} for the
$n$--point functions of the Wilson lines (= scattering amplitudes in the
eikonal approximation) and the functional JIMWLK\footnote{`JIMWLK' stands
for Jalilian-Marian, Iancu, McLerran, Weigert, Leonidov, Kovner.}
equation
\cite{JalilianMarian:1997jx,JalilianMarian:1997gr,JalilianMarian:1997dw,Kovner:2000pt,Weigert:2000gi,Iancu:2000hn,Iancu:2001ad,Ferreiro:2001qy}
for the color glass condensate (CGC = the small--$x$ part of the
wavefunction of an energetic hadron) ---, there has been little progress
towards solving these equations for correlations which are more
complicated than the 2--point function (the scattering amplitude of a
color dipole). The progress has been mostly hindered by the extreme
complexity of these equations (an issue that we shall shortly return to),
but also by the fact that, for quite some time, the dipole amplitude was
the only such a correlation to be directly relevant to phenomenology
(e.g., for deep inelastic scattering and for single--inclusive particle
production in hadron--hadron collisions; see e.g.
\cite{Iancu:2002xk,Iancu:2003xm,Gelis:2010nm} and references therein).

However, the situation has changed in the recent years, with the advent
of new, less inclusive, data, which probe higher--point functions via the
multiparticle correlations in the particle production at high energy. In
particular, the recent data at RHIC, on the azimuthal correlations in the
forward di--hadron production in deuteron--gold collisions
\cite{Braidot:2011zj}, are sensitive to a 4--point function of the Wilson
lines known as the `color quadrupole'
\cite{JalilianMarian:2004da,Nikolaev:2005dd,Baier:2005dv,Marquet:2007vb}.
These data, together with their successful description by a
phenomenological analysis using the general ideas of the CGC
\cite{Albacete:2010pg}, have revived the interest in the higher--point
correlations and triggered new efforts in that sense, mostly in relation
with the quadrupole
\cite{Dumitru:2010ak,Dominguez:2011wm,Dominguez:2011gc,Dumitru:2011vk}.
But these efforts have been hampered by the main difficulty alluded to
above --- the extreme complexity of the Balitsky--JIMWLK equations ---,
which so far has precluded the obtention of any analytic solution, even
approximate. It is our purpose in this paper to present a major step in
that sense, in the form of an approximation scheme for the higher--point
functions with $n\ge 4$, that we believe to be new. But before describing
our approach, let us briefly recall the state of the art.

The only analytic estimate for the quadrupole which is currently
available within the CGC framework is that obtained within the
McLerran--Venugopalan (MV) model \cite{McLerran:1993ni,McLerran:1993ka},
which refers to a large nucleus at not too high energy. This estimate
has first been computed in the limit of a large number of colors $N_c$ in
Ref.~\cite{JalilianMarian:2004da}, and more recently for generic $N_c$
in Ref.~\cite{Dominguez:2011wm}. The evolution equation for the
quadrupole is also known: this has first been derived in
Ref.~\cite{JalilianMarian:2004da} at large $N_c$, by using the dipole
picture \cite{Mueller:1993rr}, and then by several authors
\cite{JamalPC,GiovanniPC,Dominguez:2011gc} including ourselves, for
generic $N_c$, using the JIMWLK Hamiltonian. (Our version of the
derivation is presented in Appendix A.) But this is only one out of an
infinite hierarchy of coupled equations, which moreover exhibit a
complicated non--local structure in the transverse coordinates. This
non--locality becomes more and more severe with increasing the number $n$
of external points.

The Balitsky--JIMWLK hierarchy considerably simplifies in the
large--$N_c$ limit, where it reduces to a triangular hierarchy. It is
then enough to solve a {\em finite} number of equations in order to
determine a given correlation function. The dipole obeys a closed,
non--linear, equation, known as the Balitsky--Kovchegov (BK) equation
\cite{Balitsky:1995ub,Kovchegov:1999yj}. The quadrupole evolves according
to an inhomogeneous equation, in which the dipole plays the role of a
source term. The equation for the sextupole (a color trace of six Wilson
lines) also involves the dipole and the quadrupole, and so on. But even
in that limit, the non--locality of the equations is such that it renders
prohibitive their (numerical or analytic) study, except in the simplest
case: the BK equation, which has been studied at length
\cite{Iancu:2002xk,Iancu:2003xm,Gelis:2010nm}. As a matter of facts, the
most promising approach towards numerical studies refers to the {\em
functional} JIMWLK equation by itself, and not to the ordinary evolution
equations in the hierarchy: this functional equation can be reformulated
as a Langevin process \cite{Blaizot:2002xy}, which is well suited for the
numerics. The feasibility of such numerical solutions has been
demonstrated in Refs.~\cite{Rummukainen:2003ns,Kovchegov:2008mk,Lappi:2011ju}, which however focused on the 2--point function (the dipole amplitude) alone. It
was only very recently --- when the present study was under progress --- that this numerical method has been extended to the quadrupole \cite{Dumitru:2011vk}.

The analysis in Ref.~\cite{Dumitru:2011vk} follows the evolution of the
quadrupole with increasing energy, starting with initial conditions of
the MV type and for two special configurations of the four external
points in the transverse plane: the `line' and the `square'. (These
configurations are illustrated in Figs.~\ref{Fig:configs}.a and b below.)
Interestingly, the results show a good agreement with a heuristic
extrapolation to high energy of a formula, in particular an expression for the
quadrupole in terms of the dipole, that was derived within the MV
model \cite{JalilianMarian:2004da,Dominguez:2011wm}. Such an agreement is
hard to explain without a more fundamental understanding of the relation
between the extrapolation allude to above and the actual JIMWLK
dynamics. Moreover, the continuation of the numerical solutions in \cite{Dumitru:2011vk} to more systematic studies (say, in view of the
phenomenology) can be very difficult and demanding in computer power. Thus, it is necessary to remedy such shortcomings by completing the numerical solutions
with controlled analytic approximations --- a task that we shall accomplish here.

Specifically, our strategy will consist in solving a simplified version
of the Balitsky--JIMWLK equations at large $N_c$, in which we keep the
`virtual' terms, but we drop the `real' ones. The distinction between
`real' and `virtual' is meant here in the same sense as in the dipole
picture \cite{Mueller:1993rr}, that is, it refers to the evolution of the
projectile (dipole, quadrupole, etc): the `real' terms describe processes
in which the projectile splits via the emission of an additional, soft,
gluon (at large $N_c$, this yields e.g.~a color dipole splitting into two
dipoles), whereas the `virtual' terms refer to the complementary
probability that the projectile survive without splitting. By themselves,
the `virtual' terms control the evolution of the projectile $S$--matrix
in the vicinity of the unitarity (or `black disk') limit, so our
approximations are {\em a priori} justified in the strong scattering
regime deeply at saturation. But our scope in this paper is more general
than that: by exploiting the structure of the `virtual' terms, we shall
derive explicit expressions for the quadrupole and the sextupole in terms
of the dipole, which represent {\em global} approximations, valid for
both weak and strong scattering. We shall indeed verify that, in the
regime where the scattering is weak, our general results reduce to the
expected, linear, relations between the quadrupole, or the sextupole, and
the dipole scattering amplitude. This is not an accident: it follows from
the fact that a linear relation is always preserved by an evolution
equation derived from a Hamiltonian --- in particular the `virtual' part
of the JIMWLK Hamiltonian. (We recall here that the JIMWLK Hamiltonian is
the direct sum of the pieces responsible for the `real' and `virtual'
terms, respectively.) Hence, by using the BFKL approximation for the
dipole scattering amplitude within our general expression, we recover the
correct BFKL results for the quadrupole and the sextupole (although the
BFKL evolution is sensitive to both `real' and `virtual' terms, of
course).

Our method is general and systematic: by using similar techniques, it is
possible to derive expressions for all the $n$--point functions of the
Wilson lines in terms of the dipole $S$--matrix. Furthermore, our general
expressions are also consistent with initial conditions of the
McLerran--Venugopalan type (at large $N_c$). Hence, they provide a
unified description of the initial conditions and of their high--energy
evolution, in both the dilute (BFKL) and dense (saturation) regime. In
our opinion, the ultimate reason why such a strategy can work is the fact
that, within the JIMWLK evolution, the multi--gluon correlations are
built exclusively via high--density effects, that is, via gluon
recombination in the approach towards saturation.

The numerical analysis in Ref.~\cite{Dumitru:2011vk} provides already a
test of our approximation scheme: for the particular quadrupole
configurations investigated there (the `line' and the `square'), our
general result coincides with the high--energy extrapolation of the
respective MV formul\ae, that in Ref.~\cite{Dumitru:2011vk} were found to
agree quite well with the numerical solutions to the JIMWLK equation. The
agreement found there is even better when using the finite--$N_c$ version
of the MV formul\ae, showing that the finite--$N_c$ corrections to the
evolution of the quadrupole yield somewhat sizeable effects when $N_c=3$. This
should be contrasted with the corresponding situation for the dipole,
where one has numerically found
\cite{Rummukainen:2003ns,Kovchegov:2008mk} that the finite--$N_c$
corrections in the JIMWLK evolution are surprisingly small
--- at the level of the percent accuracy instead of the $10\%$ ($\approx
1/N_c^2$ with $N_c=3$) that would be normally expected.

Let us finally mention some possible limitations of our present analysis,
which could be improved by further studies. {\em A priori}, our
approximations are better justified for relatively symmetric
configurations of the $n$--point functions, which are such that the
transverse separations between the opposite--sign charges ({\it i.e.}
between quarks and antiquarks) are not very different from each other. It
would be interesting to test this limitation by comparing our predictions
for very asymmetric, `small--large', configurations --- which turn out to
be very interesting (see the discussion in Sect.~\ref{sect:quadru}) ---
to numerical solutions to the JIMWLK equation. We also note that, by
construction, our approximations are guaranteed to be correct in the weak
scattering regime and in the approach towards the black disk limit, but
not necessarily also in the transition between these two regimes, as
occurring around the saturation scale. Still, our results provide a
smooth (infinitely differentiable) interpolation between the two limiting
regimes and as such we believe them to be qualitatively and even
quantitatively correct including in the transition region. This is again
in agreement with the numerical analysis in Ref.~\cite{Dumitru:2011vk}.

The paper is organized as follows. In Sect.~\ref{sect:evolution} we
present the evolution equations for the dipole and the quadrupole that we
shall use as examples to illustrate general properties of the
Balitsky--JIMWLK hierarchy, like the relation between `real' and
`virtual' terms and the simplifications which appear in the large--$N_c$
limit. (The respective equations will be explicitly derived in Appendix
\ref{sect:derivation} and this will give us the opportunity to describe
the method that we shall later use, in Appendix~\ref{sect:sextupole}, to
construct the corresponding equation for the sextupole.) In
Sect.~\ref{sect:approx} we proceed with a study of the evolution deeply
as saturation, as described by the virtual terms, and introduce our
approximation scheme on a simple, pedagogical, example: the `line'
configuration for a quadrupole. In Sect.~\ref{sect:quadru} we construct
our global approximation for a generic configuration of the quadrupole
and study some limiting cases corresponding to simple, but interesting,
configurations. In particular, we shall identify configurations for which
the quadrupole is exactly factorizing in the product of two dipoles. The
corresponding analysis of the sextupole, including its evolution
equation, is given in Appendix~\ref{sect:sextupole}.

\section{Evolution equations}
\setcounter{equation}{0} \label{sect:evolution}

In this section, we shall review the  Balitsky--JIMWLK evolution
equations for the dipole and the quadrupole and explain the origin and
the physical significance of the various terms which appear in these
equations. This discussion will be useful in view of the construction of
an approximation scheme later on.

Within the CGC effective theory, valid to leading logarithmic accuracy at
high energy, the evolution of a gauge invariant operator $\hat{\mcal{O}}$
with increasing energy is determined by
 \beq\label{general}
 \frac{\del \lan \hat{\mcal{O}} \ran_Y}{\del Y}
 = \lan H \hat{\mcal{O}} \ran_Y\,,
 \eeq
where the brackets refer to the average over the color fields in the
target, as computed with the CGC weight function, and $H$ is the JIMWLK
Hamiltonian. The latter, when acting on gauge invariant observables, is
most conveniently given in \cite{Hatta:2005as} and reads
 \beq\label{H}
 H = -\frac{1}{16 \pi^3} \int_{\bmu\bmv\bmz}
 \mcal{M}_{\bmu\bmv\bmz}
 \left(1 + \wt{V}^{\dagger}_{\bmu} \wt{V}_{\bmv}
 -\wt{V}^{\dagger}_{\bmu} \wt{V}_{\bmz}
 -\wt{V}^{\dagger}_{\bmz} \wt{V}_{\bmv}\right)^{ab}
 \frac{\delta}{\delta \alpha_{\bmu}^a}
 \frac{\delta}{\delta \alpha_{\bmv}^b}.
 \eeq
In the above equations, $\alpha^a(x^-,\bmx)$ is the target color gauge
field in the covariant gauge, where $A^\mu_a =\delta^{\mu+}\alpha_a$,
$\mcal{M}$ is the dipole kernel
 \beq\label{M}
 \mcal{M}_{\bmu\bmv\bmz}\,\equiv\, \frac{(\bm{u}-\bm{v})^2}
 {(\bm{u}-\bm{z})^2(\bm{z}-\bm{v})^2}\,,\eeq
and $\wt{V}^{\dagger}$ and $\wt{V}$ are Wilson lines in the adjoint
representation:
 \beq\wt{V}^{\dagger}_{\bmx}\,\equiv\,{\mbox P}\exp \left(
\rmi g\int \rmd x^-\alpha_a(x^-,\bmx)T^a \right), \label{Vadj}
 \eeq
with P denoting path--ordering in $x^-$. Our conventions are such that
the nuclear target (the CGC) is a right--mover, whereas the projectile
(to which refers the operator $\hat{\mcal{O}}$) is a left--mover. The
functional derivative in \eqn{H} are understood to act at the largest
value of $x^-$, that is, at the upper end point of the path--ordered
exponential in \eqn{Vadj}.

The operators that we shall deal with are all built with products of
Wilson lines (in the fundamental representation, for definiteness) and
represent the $S$--matrices for systems of partons (quark and antiquarks)
scattering off the nuclear target, in the eikonal approximation.
Specifically, we shall focus on the {\em color dipole} --- a
quark--antiquark pair in an overall color singlet state, with $S$--matrix
operator
 \beq\label{Sdipole}
 \hat{S}_{\bmx_1\bmx_2} \equiv \hat{S}_{\bmx_1\bmx_2} ^{(2)}=
 \frac{1}{N_c}\,\rmtr({V}^{\dagger}_{\bmx_1} {V}_{\bmx_2})\,,
 \eeq
the {\em color quadrupole} --- a system of two quarks and two antiquarks,
for which
 \beq\label{Squadrupole}
 \hat{Q}_{\bmx_1\bmx_2\bmx_3\bmx_4} \equiv
 \hat{S}_{\bmx_1\bmx_2\bmx_3\bmx_4}^{(4)}=
 \frac{1}{N_c}\,
 \rmtr({V}^{\dagger}_{\bmx_1} {V}_{\bmx_2}{V}^{\dagger}_{\bmx_3}
 {V}_{\bmx_4})\,,
 \eeq
and the {\em color sextupole}, with
 \beq\label{Ssextupole}
 \hat{S}_{\bmx_1\bmx_2\bmx_3\bmx_4\bmx_5\bmx_6}^{(6)}=
 \frac{1}{N_c}\,
 \rmtr({V}^{\dagger}_{\bmx_1} {V}_{\bmx_2}{V}^{\dagger}_{\bmx_3}
 {V}_{\bmx_4}{V}^{\dagger}_{\bmx_5} {V}_{\bmx_6}).
 \eeq
Higher--point correlators can be similarly considered, including those
which involve the product of several color traces (see e.g.
\eqn{multitrace} below). In Eqs.~\eqref{Sdipole}--\eqref{Ssextupole},
${V}^{\dagger}$ and ${V}$ are Wilson lines in the fundamental
representation. The action of the functional derivative on these Wilson
lines reads
 \beq\label{donV}
 \frac{\delta}{\delta \alpha^a_{\bmu}}\,
 V_{\bmx}^{\dagger} =
 \rmi g \delta_{\bmx\bmu}\, t^a V_{\bmx}^{\dagger},
 \qquad
 \frac{\delta}{\delta \alpha^a_{\bmu}}\,
 V_{\bm{x}} =
 - \rmi g  \delta_{\bmx\bmu} V_{\bmx}\, t^a.
 \eeq
By using these rules within Eqs.~\eqref{general}--\eqref{H}, it is
straightforward to derive the evolution equations satisfied by the
expectation values of the three operators introduced above. A streamlined
derivation of the respective equations for the dipole and the quadrupole
will presented in Appendix~\ref{sect:derivation}, while the corresponding
equation for the sextupole will be shown (without the details of the
derivation) in Appendix~\ref{sect:sextupole}.

The ensuing equation for the dipole looks relatively simple (we denote
$\abar\equiv\alpha_s N_c/\pi$):
 \beq\label{BK}
 \frac{\del \lan \hat{S}_{\bmx_1\bmx_2} \ran_Y}{\del Y}=
 \frac{\abar}{2\pi}\, \int_{\bmz}
 \mcal{M}_{\bmx_1\bmx_2\bmz}
 \lan \hat{S}_{\bmx_1\bmz} \hat{S}_{\bmz\bmx_2}
 -\hat{S}_{\bmx_1\bmx_2} \ran_Y\,,
 \eeq
but that for the quadrupole is considerably more involved:
 \beq\label{Qevol}
 \frac{\del \lan\hat{Q}_{\bmx_1\bmx_2\bmx_3\bmx_4} \ran_Y}{\del Y} =
 \frac{\abar}{4\pi}  \int_{\bmz}\!\!\!\!\!\! &{}&
 \Big[(\mcal{M}_{\bmx_1\bmx_2\bmz} +
 \mcal{M}_{\bmx_1\bmx_4\bmz} -
 \mcal{M}_{\bmx_2\bmx_4\bmz})
 \lan
 \hat{S}_{\bmx_1\bmz}\hat{Q}_{\bmz\bmx_2\bmx_3\bmx_4}
 \ran_Y
 \nn
 &+&(\mcal{M}_{\bmx_1\bmx_2\bmz} +
 \mcal{M}_{\bmx_2\bmx_3\bmz} -
 \mcal{M}_{\bmx_1\bmx_3\bmz})
 \lan
 \hat{S}_{\bmz\bmx_2}\hat{Q}_{\bmx_1\bmz\bmx_3\bmx_4}
 \ran_Y
 \nn
 &+&(\mcal{M}_{\bmx_2\bmx_3\bmz} +
 \mcal{M}_{\bmx_3\bmx_4\bmz} -
 \mcal{M}_{\bmx_2\bmx_4\bmz})
 \lan
 \hat{S}_{\bmx_3\bmz}\hat{Q}_{\bmx_1\bmx_2\bmz\bmx_4}
 \ran_Y
 \nn
 &+&(\mcal{M}_{\bmx_1\bmx_4\bmz} +
 \mcal{M}_{\bmx_3\bmx_4\bmz} -
 \mcal{M}_{\bmx_1\bmx_3\bmz})
 \lan
 \hat{S}_{\bmz\bmx_4}\hat{Q}_{\bmx_1\bmx_2\bmx_3\bmz}
 \ran_Y
 \nn
 &-&(\mcal{M}_{\bmx_1\bmx_2\bmz} + \mcal{M}_{\bmx_3\bmx_4\bmz}
 +\mcal{M}_{\bmx_1\bmx_4\bmz} + \mcal{M}_{\bmx_2\bmx_3\bmz})
 \lan
 \hat{Q}_{\bmx_1\bmx_2\bmx_3\bmx_4}
 \ran_Y
 \nn
 &-&(\mcal{M}_{\bmx_1\bmx_2\bmz} + \mcal{M}_{\bmx_3\bmx_4\bmz}
 -\mcal{M}_{\bmx_1\bmx_3\bmz} - \mcal{M}_{\bmx_2\bmx_4\bmz})
 \lan
 \hat{S}_{\bmx_1\bmx_2}\hat{S}_{\bmx_3\bmx_4}
 \ran_Y
 \nn
 &-&(\mcal{M}_{\bmx_1\bmx_4\bmz} + \mcal{M}_{\bmx_2\bmx_3\bmz}
 -\mcal{M}_{\bmx_1\bmx_3\bmz} - \mcal{M}_{\bmx_2\bmx_4\bmz})
 \lan
 \hat{S}_{\bmx_3\bmx_2}\hat{S}_{\bmx_1\bmx_4}
 \ran_Y\Big].
 \eeq
We shall now discuss the role of the various terms in the above
equations, in order to explain, in particular, the difference between
`real' and `virtual' terms.

Consider first \eqn{BK} for the dipole $S$--matrix. The `real' term is
the first term in the right hand side, which is quadratic in $\hat{S}$.
When interpreted in terms of projectile evolution, this term describes
the splitting of the original dipole $(\bmx_1,\,\bmx_2)$ into two new
dipoles $(\bmx_1,\,\bmz)$ and $(\bmz,\,\bmx_2)$, which then scatter off
the target. The `virtual' term, that is, the negative term which is
linear in $\hat{S}$, describes the reduction in the probability that the
dipole survive in its original state --- that is, the probability for the
dipole not to split.  The relative size of these two types of terms
(`real' and `virtual') is constrained by probability conservation, as
correctly encoded in the JIMWLK Hamiltonian. Remarkably, the `real' term
has been fully generated by the {\em last} two terms in the Hamiltonian
\eqref{H}, whereas the `virtual' term originated from the {\em first} two
terms there. All the terms appearing in \eqn{BK} are of leading order in
$N_c$. Subleading terms, of relative order $1/N_c^2$, had been also
generated, at intermediate steps, by each of the four terms in the
Hamiltonian, but they have exactly canceled after summing up all the
contributions. Note the cancelation of `ultraviolet' ({\em i.e.}
short--distance) singularities between the `real' and the `virtual'
terms: the dipole kernel \eqref{M} has poles at $\bm{z}=\bm{u}$ and
$\bm{z}=\bm{v}$, which within the integral over $\bm{z}$ generate
logarithmic divergences, separately for the `real' and for the `virtual'
term. However, these divergences cancel in the overall equation, because
of probability conservation together with the property of `color
transparency' : $\hat{S}_{\bmx_1\bmx_2}\to 1$ when $\bmx_1\to\bmx_2$. The
latter is merely the statement that a `color dipole' of zero transverse
size is a totally neutral object, which cannot interact.

An entirely similar discussion applies to the evolution equation
\eqref{Qevol} for the quadrupole. The terms involving $\lan \hat{S}
\hat{Q}\ran_Y$ in the right hand side are `real' terms describing the
splitting of the original quadrupole into a new quadrupole plus a dipole,
and have been all generated by the action of the last two terms in the
Hamiltonian \eqref{H}. The `virtual' terms involving $\lan \hat{Q}\ran_Y$
and $\lan \hat{S} \hat{S}\ran_Y$ are necessary for probability
conservation, and have been generated by the first two terms in the
Hamiltonian. Once again, all the terms subleading at large $N_c$ (as
generated by the individual pieces of the Hamiltonian) have canceled in
the final equation. Furthermore, all possible ultraviolet divergencies
due to the poles of the dipole kernel cancel between  the `real' and the
`virtual' terms, due to color transparency.

All the  above features are generic: they also apply to the
Balitsky--JIMWLK evolution of the sextupole and, more generally, to any
(gauge--invariant) operator which involves a single trace of products of
pairs of Wilson lines, of the form
 \beq\label{S2n}
 \hat{S}_{\bmx_1\bmx_2 ...\bmx_{2n\!-\!1}\bmx_{2n}}^{(2n)} =\,
 \frac{1}{N_c}\,
 \rmtr({V}^{\dagger}_{\bmx_1} {V}_{\bmx_2}...
 {V}^{\dagger}_{\bmx_{2n\!-\!1}}{V}_{\bmx_{2n}}).
 \eeq
As for the multi--trace operators, of the form
 \beq\label{multitrace}
 \hat{\mcal{O}} = \,\frac{1}{N_c}\,\rmtr({V}^{\dagger}_{\bmx_1} {V}_{\bmx_2}...)
 \frac{1}{N_c}\,\rmtr({V}^{\dagger}_{\bmy_1} {V}_{\bmy_2}...)
 \frac{1}{N_c}\,\rmtr({V}^{\dagger}_{\bmz_1} {V}_{\bmz_2}...),
 \eeq
the only new feature is that the respective evolution equations involve
genuine $1/N_c^2$ corrections, as generated when the two functional
derivatives in \eqn{H} act on Wilson lines which belong to different
traces (see e.g. Appendix F in \cite{Triantafyllopoulos:2005cn} for an
example).

As manifest by inspection of Eqs.~\eqref{BK} and \eqref{Qevol}, these
equations are generally not closed: e.g., the equation for the quadrupole
also involves the 4--point function  $\lan \hat{S} \hat{S}\ran_Y$ and the
6--point function $\lan \hat{S} \hat{Q}\ran_Y$, which in turn are coupled
(via the respective evolution equations) to even higher--point
correlators, so altogether one has to deal with an infinite hierarchy of
equations. But important simplifications occur in the large--$N_c$ limit,
as anticipated in the Introduction. Then, for a multi--trace operator
like \eqref{multitrace}, it is not hard to show that the hierarchy admits
the factorized solution
 \beq\label{fact}
 \lan\hat{\mcal{O}}\ran_Y =
 \Big\lan\frac{1}{N_c}\,\rmtr({V}^{\dagger}_{\bmx_1} {V}_{\bmx_2}...)\Big\ran_Y
 \Big\lan\frac{1}{N_c}\,\rmtr({V}^{\dagger}_{\bmy_1} {V}_{\bmy_2}...)\Big\ran_Y
 \Big\lan\frac{1}{N_c}\,\rmtr({V}^{\dagger}_{\bmz_1} {V}_{\bmz_2}...)\Big\ran_Y...,
 \eeq
provided this factorization is already satisfied by the initial
conditions. Then the hierarchy becomes triangular and therefore finite:
\eqn{BK} becomes a closed equation for $\lan\hat{S}\ran_Y$ (the BK
equation), while \eqn{Qevol} becomes an inhomogeneous equation for $\lan
\hat{Q} \ran_Y$ with coefficients which depend upon $\lan\hat{S}\ran_Y$.
This is the equation originally derived in \cite{JalilianMarian:2004da}
and that we shall further study in the next section.

\section{From virtual terms to global approximations}
\setcounter{equation}{0} \label{sect:approx}

In order to better appreciate the approximation scheme that we shall
eventually propose, it is useful to investigate the behavior of the
solutions to the evolution equations introduced in the previous section
in two limiting regimes: the dilute, or weak scattering, regime, where
all the transverse separations $r_{ij} \equiv |\bmx_i - \bmx_j|$ between
the external points are very small compared to $1/Q_s(Y)$, and the dense,
or strong scattering, regime, where all these distances are much larger
than $1/Q_s(Y)$. Here, $Q_s(Y)\propto \rme^{\omega\abar Y}$ is the
saturation momentum in the target, and it is also the transverse momentum
scale which marks the transition from weak to strong scattering, for both
the dipole and the quadrupole.

The discussion of the dilute regime is straightforward. Introducing the
dipole $T$--matrix operator $\hat{T}_{\bmx_1\bmx_2} \equiv 1 -
\hat{S}_{\bmx_1\bmx_2}$ and the corresponding scattering amplitude $\lan
\hat{T}_{\bmx_1\bmx_2} \ran_Y$, then for $r  = |\bmx_1-\bmx_2|\ll 1/Q_s$
the scattering is weak, $\lan\hat{T}_{\bmx_1\bmx_2}\ran_Y \ll 1$, and the
amplitude obeys the linearized (in $\lan \hat{T} \ran_Y$) version of
\eqn{BK}, that is, the BFKL equation:
 \beq\label{BFKL}
 \frac{\del \lan \hat{T}_{\bmx_1\bmx_2} \ran_Y}{\del Y}=
 \atpi\, \int_{\bmz}
 \mcal{M}_{\bmx_1\bmx_2\bmz}
 \lan \hat{T}_{\bmx_1\bmz} + \hat{T}_{\bmz\bmx_2}
 -\hat{T}_{\bmx_1\bmx_2} \ran_Y\,.
 \eeq
Consider similarly the quadrupole: when $r_{ij}\ll 1/Q_s$ for all the six
pairs $(i,\,j)$ (with $i,\,j=1,2,3,4$) of external points, the scattering
is necessarily weak, because there is no net color charge in the
projectile over an area $\sim 1/Q_s^2$ (which is the area where high
gluon density effects become important in the target). Then, for
consistency, it is enough to keep the lowest order terms, of order
$g^2\alpha^2$, in the perturbative expansion of $1-\lan\hat{Q} \ran_Y$.
These are obtained by expanding the Wilson lines in \eqn{Squadrupole} up
to quadratic order. After also comparing with the corresponding expansion
of $\hat{S}$ in Eq.~(\ref{Sdipole}), one finds
 \beq\label{QBFKL}
1 -\lan \hat{Q}_{\bmx_1\bmx_2\bmx_3\bmx_4}  \ran_Y\simeq
  \lan\hat{T}_{\bmx_1\bmx_2}
 -\hat{T}_{\bmx_1\bmx_3}
 +\hat{T}_{\bmx_1\bmx_4}
 +\hat{T}_{\bmx_2\bmx_3}
 -\hat{T}_{\bmx_2\bmx_4}
 +\hat{T}_{\bmx_3\bmx_4} \ran_Y,
 \eeq
where it is understood that, in the r.h.s., $\lan \hat{T} \ran_Y$ is the
dipole amplitude in the dilute regime and obeys the BFKL equation
\eqref{BFKL}. \eqn{QBFKL} also tells us that, for generic configurations
at least, the quadrupole scattering can become strong when at least one
(which necessarily means at least three) of the six transverse distances
$r_{ij}$ is of order $1/Q_s$, or larger. However, care must be taken when
discussing very asymmetric configurations (see below).

Consider now the strong scattering regime at $r_{ij}\gg 1/Q_s(Y)$, where
we shall study the approach towards the `black disk' limit, {\em i.e.}
the way how $S$--matrices like $\lan \hat{S} \ran_Y$ and $\lan \hat{Q}
\ran_Y$ approach to zero. The corresponding analysis for the dipole is
well known \cite{Levin:1999mw,Iancu:2003zr}, but it will be briefly
reviewed here, in preparation for the quadrupole and higher $n$--point
functions. The crucial observation is that, at least for large $N_c$,
this approach is controlled by the virtual terms in the respective
evolution equations\footnote{This statement is strictly correct only
within the limits of the JIMWLK evolution. As emphasized in
\cite{Iancu:2003zr}, the inclusion of Pomeron loops
\cite{Iancu:2004iy,Mueller:2005ut,Iancu:2005nj,Kovner:2005nq} in the
analysis would modify the approach towards the black disk limit, even for
large $N_c$. Still, the functional form of this approach, as shown in
\eqn{SLT}, should remain correct: only the overall coefficient in front
of the logarithm squared should be reduced by a factor of 2
\cite{Iancu:2003zr}.}. Indeed, at large $N_c$ we have, e.g., $\lan
\hat{S}\hat{S} \ran_Y \simeq \lan\hat{S}\ran_Y \lan \hat{S} \ran_Y$, and
then the `real' term in the r.h.s. of \eqn{BK}, which is quadratic in
$\lan \hat{S} \ran_Y$, is much smaller than the linear, `virtual', term
there whenever $\lan \hat{S} \ran_Y\ll 1$. More precisely, when $r  =
|\bmx_1-\bmx_2|\gg 1/Q_s(Y)$, the linear term dominates over the
quadratic one unless either $|\bmx_1 - \bmz|$ or $|\bmz - \bmx_2|$ is as
small as $1/Q_s(Y)$. Then, the r.h.s. of \eqn{BK} is dominated by the
logarithmic regions of integration where one has
  \beq\label{conds}
 1/Q_s(Y) \ll |\bmx_1 - \bmz|,\,|\bmz - \bmx_2| \ll r\,.
 \eeq
This discussion implies that the approach of the dipole $S$--matrix towards
the `black disk' limit is determined by the following, simplified,
equation
  \beq\label{BKstrong}
 \frac{\del \ln\lan\hat{S}(r)\ran_Y}{\del Y}
 \simeq -\abar \int_{1/Q_s^2}^{r^2}\frac{\dif z^2}{z^2}
 = -\abar \ln [r^2 Q_s^2(Y)],
 \eeq
which is correct to leading logarithmic accuracy with respect to the
large logarithm $\ln (r^2 Q_s^2)$. That is, in writing the r.h.s. of
\eqn{BKstrong}, we control the coefficient in front of the logarithm but
not also the constant term under the logarithm. Using the fixed coupling
relation $\ln [r^2 Q_s^2(Y)] \simeq \omega\abar (Y-Y_0)$, where $Y_0$ is
such that $Q_s(Y_0)=1/r$, we finally arrive at
 \beq\label{SLT}
 \lan \hat{S}(r) \ran_Y \,\simeq\,
 \lan \hat{S}(r) \ran_{Y_0}
 \,
 \exp\bigg\{ - \frac{1}{2\omega}\ln^2 [r^2 Q_s^2(Y)] \bigg\}.
 \eeq
This expression is valid for $Y\ge Y_0$, where $Y_0$ is roughly the
rapidity at which the scattering of the dipole with size $r$ becomes
strong.

We now turn to the quadrupole and similarly study the strong scattering
regime for generic configurations of the four external points in the
transverse plane. Note however that our ultimate objective in this study
is different, and actually more ambitious, than in the corresponding
study of the dipole: what we are truly interested in, is not the law for
the approach towards the black disk limit (although this law will emerge
too from our subsequent analysis), but rather the functional relation
between the quadrupole $S$--matrix $\lan\hat{Q} \ran_Y$ and the dipole
one $\lan\hat{S} \ran_Y$, as predicted by the solutions to the respective
evolution equations with the virtual terms alone. Indeed, as announced in
the Introduction and will be explicitly checked on the final results,
this functional form is correct also outside the strong scattering regime
--- namely, it has the right limit at weak scattering, as shown in
\eqn{QBFKL}.

Note first that a quadrupole configuration is bound to be in the strong
scattering regime whenever all the four quark--antiquark separations
($r_{12}$,  $r_{14}$, $r_{23}$ and $r_{34}$) are much larger than
$1/Q_s$, independently of the relative separations between the two quarks
($r_{13}$) and the two antiquarks ($r_{24}$). Indeed, when $r_{q\bar
q}\gg 1/Q_s$, there is uncompensated color charge over distances $\gtrsim
1/Q_s$, which implies that the scattering is strong. In this regime and
for large $N_c$, the evolution towards the unitarity limit is controlled
by the virtual terms, by the same arguments as for the dipole. After
neglecting the `real' terms, \eqn{Qevol} simplifies considerably and, in
particular, it becomes `diagonal' with respect to the quadrupole
configuration: both the l.h.s. and the r.h.s. involve the same
configuration $(\bmx_1,\bmx_2,\bmx_3,\bmx_4)$. This makes it possible to
study the approach towards the unitarity limit configuration by
configuration and thus obtain explicit, analytic solutions.

\begin{figure}
\begin{minipage}[b]{0.245\textwidth}
\begin{center}
\includegraphics[scale=0.75]{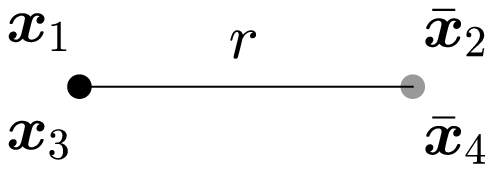}\\(a)
\end{center}
\end{minipage}
\begin{minipage}[b]{0.245\textwidth}
\begin{center}
\includegraphics[scale=0.75]{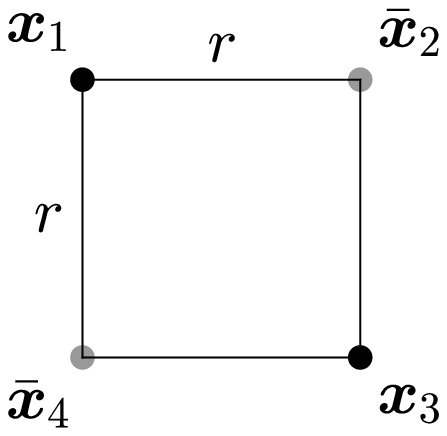}\\(b)
\end{center}
\end{minipage}
\begin{minipage}[b]{0.245\textwidth}
\begin{center}
\includegraphics[scale=0.75]{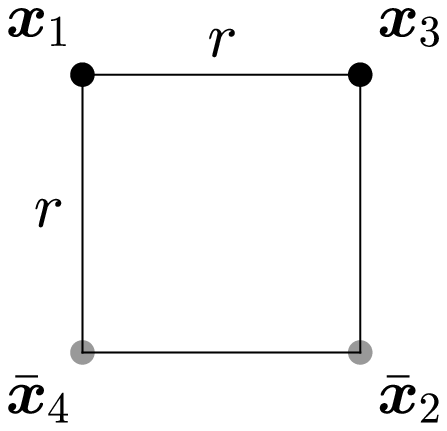}\\(c)
\end{center}
\end{minipage}
\begin{minipage}[b]{0.245\textwidth}
\begin{center}
\includegraphics[scale=0.75]{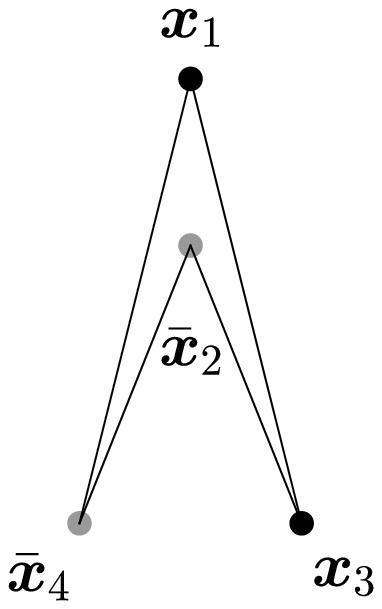}\\(d)
\end{center}
\end{minipage}
\caption{\sl Special quadrupole configurations
that we shall use to illustrate our general results.
The dots represent the positions of the fermions in the transverse
plane. (A bar on a transverse coordinate refers to an antiquark.)
The lines have no physical meaning, they are meant only to delimitate
the shape of the figures, in conformity with the names that we use
to refer to them.}
\label{Fig:configs}
\end{figure}

The case of a general configuration (which respects the conditions stated
above) will be addressed in the next section. But before that general
discussion, it is instructive to consider a simple, particular case as a
warm--up. In what follows, we shall often use the four special
configurations shown in Fig.~\ref{Fig:configs} in order to illustrate our
results. These configurations are useful in that they involve only few
independent transverse separations $r_{ij}$, due to their high degree of
symmetry. Besides, two of them --- the `line' in Fig.~\ref{Fig:configs}.a
and the first `square' in Fig.~\ref{Fig:configs}.b --- have been
previously chosen in the numerical studies in \cite{Dumitru:2011vk}, so
for them we know already the respective numerical predictions of the
JIMWLK equation. This information will be useful for testing our
forthcoming analytic results.

We shall pick the `line' configuration, where we place the two quarks on
top of each other and the same for the antiquarks (see
Fig.~\ref{Fig:configs}.a), as our warm--up example. For this example, we
shall perform in detail all the steps to be repeated for the general case
in the next section. (The three other configurations in
Fig.~\ref{Fig:configs} will be studied in Sect.~\ref{sect:quadru}, as
special limits of our general result for the quadrupole.) As obvious from
Fig.~\ref{Fig:configs}.a, the `line' configuration involves a single
transverse scale $r$, which is the common size of the four $q\bar q$
pairs: $r_{12}=r_{14}=r_{23}=r_{34}\equiv r$. When applied to this
configuration, the general evolution equation \eqref{Qevol} reduces to
 \beq\label{Qline}
 \frac{\del \lan \hat{Q}_{\bmx_1\bmx_2\bmx_1\bmx_2}\ran_Y}{\del Y}
 &=& \frac{\abar}{\pi}
 \int_{\bmz}
 \mcal{M}_{\bmx_1\bmx_2\bmz}
 \big[\lan \hat{S}_{\bmx_1\bmz} \ran_Y
\lan\hat{Q}_{\bmz\bmx_2\bmx_1\bmx_2}\ran_Y
 +\lan \hat{S}_{\bmz\bmx_2} \ran_Y \lan \hat{Q}_{\bmx_1\bmz\bmx_1\bmx_2} \ran_Y
 \nn
 &{}&\qquad \qquad \qquad- \lan\hat{Q}_{\bmx_1\bmx_2\bmx_1\bmx_2}\ran_Y
 -\lan\hat{S}_{\bmx_1\bmx_2}\ran_Y^2
 \big].
 \eeq
When the separation $r$ is much larger than $1/Q_s$, the virtual terms
(the last two terms in \eqn{Qline}) dominate and give
 \beq\label{Qstrong2}
 \frac{\del \lan \hat{Q}(r)\ran_Y}{\del Y} \simeq
 - 2 \abar \ln (r^2 Q_s^2) \,
 \big[\lan \hat{Q}(r)\ran_Y + \lan\hat{S}(r)\ran_Y^2\big]\,.
 \eeq
We have written the argument of the quadrupole simply as $\hat{Q}(r)$,
because $r$ is the only independent transverse scale for this particular
configuration. After also using \eqn{BKstrong}, it turns out that this
inhomogeneous equation admits a relatively simple solution, which is (for
$Y \ge Y_0$)
 \beq\label{QlineS}
\lan \hat{Q}(r)\ran_Y = \lan \hat{S}(r)\ran_Y^2 \left[\frac{\lan
\hat{Q}(r)\ran_{Y_0}}{\lan \hat{S}(r)\ran_{Y_0}^2} + \ln \frac{\lan
\hat{S}(r)\ran_Y^2} {\lan \hat{S}(r)\ran_{Y_0}^2}
\right]\qquad\mbox{(`line' in Fig.~\ref{Fig:configs}.a)}\,.
 \eeq
In view of its derivation above, one may expect this formula to hold only
deeply at saturation, that is, for $Y > Y_0$, with $Y_0$ the rapidity at
which saturation sets in on the given transverse scale $r$. But as a
matter of facts, \eqn{QlineS} is more general than that: it has also the
right limit at {\em weak} scattering, in agreement with \eqn{QBFKL}, and
hence it provides a {\em global} approximation. To see that, let us study
the weak scattering limit of \eqn{QlineS}, by writing $\lan
\hat{S}(r)\ran_Y=1 - \lan \hat{T}(r)\ran_Y$, with $\lan
\hat{T}(r)\ran_Y\ll 1$, and then expanding the r.h.s. to linear order in
$\lan \hat{T}(r)\ran_Y$. After doing that, one finds that the following
relation is satisfied at $Y$, provided it was already satisfied at $Y_0$
:
  \beq\label{lBFKL}
  \lan \hat{Q}(r)\ran_Y \simeq 1- 4\lan \hat{T}(r)\ran_Y\,.\eeq
As anticipated, \eqn{lBFKL} is the correct weak--scattering limit for
this particular configuration, as predicted by \eqn{QBFKL}.

\eqn{QlineS} can be further simplified if one assumes that the initial
conditions at $Y_0$ are provided by the MV model at large $N_c$. Indeed,
it is easy to check that \eqn{QlineS} reduces to
 \beq\label{QL}
 \lan \hat{Q}(r)\ran_Y =
 \lan \hat{S}(r)\ran_Y^2
 \left[1 +
 \ln {\lan \hat{S}(r)\ran_Y^2}
 \right]\,,
 \eeq
provided the above formula already holds at the initial rapidity $Y_0$
--- a property which is indeed satisfied within the MV model at large
$N_c$.

To summarize, by using \eqn{QL} together with a reasonable approximation
for the dipole $S$--matrix $\lan \hat{S}(r)\ran_Y$ --- say, the solution
to the BK equation with initial conditions provided by the large--$N_c$
version of the MV model --- one obtains a unified description for
\texttt{(i)} the initial conditions at low energy, \texttt{(ii)} the BFKL
regime at high energy, and \texttt{(iii)} the approach towards the black
disk limit at all the energies.

As demonstrated by the numerical analysis in Ref.~\cite{Dumitru:2011vk},
\eqn{QL} agrees well with the numerical solution to the JIMWLK
equation for this particular configuration. In the forthcoming section
and in Appendix \ref{sect:sextupole}, we shall generalize \eqn{QlineS} to
generic configurations for the quadrupole and the sextupole, and we shall
also study some other particular configurations.

\section{A global approximation for the quadrupole}
\setcounter{equation}{0} \label{sect:quadru}

We shall now extend the analysis in the previous section to a generic
configuration  for the quadrupole --- that is, we shall study the
approximation obtained by keeping only the virtual terms in \eqn{Qevol}.
In line with our general strategy, we shall rely on this approximation,
which is strictly speaking valid only in the vicinity of the black disk
limit, in order to derive a functional relation between the quadrupole
and the dipole, that will be then promoted to a global approximation.

After neglecting the `real' terms and restricting the integrations over
$\bmz$ to the regions yielding large logarithmic contributions, like in
\eqn{conds}, we get
 \beq\label{Qevolvirtual}
 \hspace*{-.4cm}
 \frac{\del \lan\hat{Q}_{\bmx_1\bmx_2\bmx_3\bmx_4}
 \ran_Y}{\del Y} \simeq\,
 &-&\frac{\abar}{2}
 \left[ \ln(r_{12}^2 Q_s^2) + \ln(r_{34}^2 Q_s^2)
 + \ln(r_{14}^2 Q_s^2) + \ln(r_{23}^2 Q_s^2) \right]
 \lan \hat{Q}_{\bmx_1\bmx_2\bmx_3\bmx_4} \ran_Y
 \nn
 &-&\frac{\abar}{2}
 \left[ \ln(r_{12}^2 Q_s^2) + \ln(r_{34}^2 Q_s^2)
 - \ln(r_{13}^2 Q_s^2) - \ln(r_{24}^2 Q_s^2) \right]
 \lan \hat{S}_{\bmx_1\bmx_2}\ran_Y\lan\hat{S}_{\bmx_3\bmx_4} \ran_Y
 \nn
 &-&\frac{\abar}{2}
 \left[ \ln(r_{14}^2 Q_s^2) + \ln(r_{23}^2 Q_s^2)
 - \ln(r_{13}^2 Q_s^2) - \ln(r_{24}^2 Q_s^2) \right]
 \lan \hat{S}_{\bmx_3\bmx_2}\ran_Y\lan\hat{S}_{\bmx_1\bmx_4} \ran_Y,
 \nn
 \eeq
where in writing the r.h.s. we have also factorized the 2--dipole
$S$--matrix, as appropriate at large $N_c$. Now we can use \eqn{BKstrong}
in order to express the logarithms in terms of the dipole $S$-matrix. The
merit of that is three-fold. First, we can express the quadrupole in
terms of the dipole without worrying about the explicit solution for the
latter. Second, in order to get the dominant logarithmic contribution we
do the same kind of approximation in both the dipole and the quadrupole
equations since their structures are the same. Therefore we probably have
better control than originally announced, that is, we have better
accuracy than the one given by the leading logarithmic terms. Third, we
do not really need to specify the energy dependence of $Q_s$ and thus the
approach is fully valid even if we consider a running coupling scenario
(at least in the case where $\abar$ is evaluated at $Q_s$). Thus, writing
all logarithms in \eqn{Qevolvirtual} terms of $S$ we have
 \beq\label{QevolS}
 \frac{\del \lan\hat{Q}_{\bmx_1\bmx_2\bmx_3\bmx_4} \ran_Y}{\del Y} &=&
 \frac{1}{2}
 \left[\frac{\del}{\del Y}
 \ln \lan \hat{S}_{\bmx_1\bmx_2} \ran_Y
 \lan \hat{S}_{\bmx_3\bmx_2} \ran_Y
 \lan \hat{S}_{\bmx_3\bmx_4} \ran_Y
 \lan \hat{S}_{\bmx_1\bmx_4} \ran_Y
 \right]
 \lan\hat{Q}_{\bmx_1\bmx_2\bmx_3\bmx_4} \ran_Y
 \nn
 &+&
 \frac{1}{2}
 \left[\frac{\del}{\del Y}
 \ln \frac{\lan \hat{S}_{\bmx_1\bmx_2} \ran_Y
 \lan \hat{S}_{\bmx_3\bmx_4} \ran_Y}
 {\lan \hat{S}_{\bmx_1\bmx_3} \ran_Y
 \lan \hat{S}_{\bmx_2\bmx_4} \ran_Y}
 \right]
 \lan \hat{S}_{\bmx_1\bmx_2} \ran_Y
 \lan \hat{S}_{\bmx_3\bmx_4} \ran_Y
 \nn
 &+&
 \frac{1}{2}
 \left[\frac{\del}{\del Y}
 \ln \frac{\lan \hat{S}_{\bmx_1\bmx_4} \ran_Y
 \lan \hat{S}_{\bmx_3\bmx_2} \ran_Y}
 {\lan \hat{S}_{\bmx_1\bmx_3} \ran_Y
 \lan \hat{S}_{\bmx_2\bmx_4} \ran_Y}
 \right]
 \lan \hat{S}_{\bmx_1\bmx_4} \ran_Y
 \lan \hat{S}_{\bmx_3\bmx_2} \ran_Y.
 \eeq
This is an ordinary first order inhomogeneous differential equation whose
general solution is straightforward to find; it reads
 \beq\label{QsolS}
 &{}&\lan\hat{Q}_{\bmx_1\bmx_2\bmx_3\bmx_4} \ran_Y =
 \sqrt{\lan \hat{S}_{\bmx_1\bmx_2} \ran_Y
 \lan \hat{S}_{\bmx_3\bmx_2} \ran_Y
 \lan \hat{S}_{\bmx_3\bmx_4} \ran_Y
 \lan \hat{S}_{\bmx_1\bmx_4} \ran_Y
 }
 \left[
 \frac{\lan\hat{Q}_{\bmx_1\bmx_2\bmx_3\bmx_4} \ran_{Y_0}}
 {\sqrt{\lan \hat{S}_{\bmx_1\bmx_2} \ran_{Y_0}
 \lan \hat{S}_{\bmx_3\bmx_2} \ran_{Y_0}
 \lan \hat{S}_{\bmx_3\bmx_4} \ran_{Y_0}
 \lan \hat{S}_{\bmx_1\bmx_4} \ran_{Y_0}
 }}\right.
 \nn
 &{}& \quad\left.+\, \frac{1}{2}\,\int_{Y_0}^{Y} \dif y\,
 \frac{\lan \hat{S}_{\bmx_1\bmx_3} \ran_y
 \lan \hat{S}_{\bmx_2\bmx_4} \ran_y}
 {\sqrt{\lan \hat{S}_{\bmx_1\bmx_2} \ran_y
 \lan \hat{S}_{\bmx_3\bmx_2} \ran_y
 \lan \hat{S}_{\bmx_3\bmx_4} \ran_y
 \lan \hat{S}_{\bmx_1\bmx_4} \ran_y
 }}\,
 \frac{\del}{\del y}
 \frac{\lan \hat{S}_{\bmx_1\bmx_2} \ran_y
 \lan \hat{S}_{\bmx_3\bmx_4} \ran_y+
 \lan \hat{S}_{\bmx_1\bmx_4} \ran_y
 \lan \hat{S}_{\bmx_3\bmx_2} \ran_y}{\lan \hat{S}_{\bmx_1\bmx_3} \ran_y
 \lan \hat{S}_{\bmx_2\bmx_4} \ran_y}\right]\!\!.
 \eeq
As explained in the Introduction, the above expression is guaranteed to
be correct also for weak scattering, even though it has been obtained
here via manipulations which are well justified only deeply at
saturation. This is so since, at weak scattering, the relation between
the quadrupole and the dipole becomes linear, cf. \eqn{QBFKL}. Such a
linear relation is then preserved by any evolution equation generated by
a Hamiltonian --- in particular, by the equations including only the
virtual terms, which are generated, as we have seen, by a truncated
version of the JIMWLK Hamiltonian (the first two terms in \eqn{H}). This
argument about linearity is simple, but at the same time important for
the present construction, so let us make a small digression in order to
explain it better.

Consider a set of three operators, $\hat A$, $\hat B$, and $\hat C$,
which obey a linear relation: $\hat A= c_1 \hat B + c_2 \hat C$. Also
assume that the evolution of these operators with `time' $Y$ is governed
by some Hamiltonian $H$; e.g. $\partial_Y \hat A=H\hat A$. Then one can
successively write
 \beq
 \frac{\del \hat A}{\del Y}\,=\,
 H\hat A\,=\,H\big(c_1\hat B+c_2\hat C\big)\,=\,
 c_1\frac{\del \hat B}{\del Y}\,+
 \,c_2\frac{\del \hat C}{\del Y}\,.
 \eeq
This equation implies that the linear relation $\hat A= c_1 \hat B + c_2
\hat C$ is preserved at any $Y$ provided it was satisfied at the initial
`time' $Y_0$. Note that specific form of $H$ was irrelevant for the above
argument; all that matters is the fact that this is a linear operator.

Returning to the physics problem at hand, the relation \eqref{QsolS}
between the quadrupole and the dipole has been obtained by solving a
Hamiltonian evolution equation. The corresponding Hamiltonian is
incorrect in the dilute regime, where it would predict a wrong evolution
for the quadrupole, or the dipole, {\em taken separately}. However, as a
linear operator, it must preserve the linear relation \eqref{QBFKL}
between the two operators $\hat{Q}$ and $\hat{S}$ which is valid in that
regime. And indeed, it can be easily checked --- by rewriting $\lan
\hat{S}_{\bmx_i\bmx_j} \ran_y=1- \lan \hat{T}_{\bmx_i\bmx_j} \ran_y$ and
then expanding the r.h.s. of \eqn{QsolS} to linear order in the various
$\lan \hat{T}_{\bmx_i\bmx_j} \ran_y$'s --- that \eqn{QsolS} predicts the
correct relation, \eqn{QBFKL}, between the quadrupole and the dipole at
weak scattering and at rapidity $Y$ provided this relation was already
satisfied at the initial rapidity $Y_0$.

In view of the above, we conclude that \eqn{QsolS}, when used with a
numerical solution of the BK equation (or some good approximation to it),
provides a reliable approximation to the solution to the quadrupole
equation which is valid in all regimes (at least for generic
configurations, which are not very asymmetric). This solution provides a
smooth (infinitely differentiable) interpolation between the BFKL
solution in the weak scattering regime at $r_{ij}\ll 1/Q_s$ and the
correct approach towards the black disk limit at $r_{ij}\gg 1/Q_s$.

The expression \eqref{QsolS} has some other good properties. It is
symmetric under the interchange of the two quarks (or the two antiquarks)
at any $Y$, provided that this is true at the initial rapidity $Y_0$.
This is a true property of the JIMWLK evolution at large-$N_c$, as it can be directly checked at the level of the evolution equations in
Sect.~\ref{sect:evolution}. Furthermore, there is an interesting relation
between  \eqn{QsolS} and the corresponding expression in the MV model
\cite{JalilianMarian:2004da,Dominguez:2011wm}. To see this connection, it
is useful to write the dipole $S$--matrix in the form
 \beq\label{Sgaussian}
 \lan \hat{S}_{\bmx_i\bmx_j}\ran_Y =
 \rme^{-\Gamma_Y(\bmx_i,\bmx_j)}
 \lan \hat{S}_{\bmx_i\bmx_j}\ran_{Y_0}.
 \eeq
Now, \eqref{QsolS} becomes formally identical with the quadrupole formula
in the MV model  (for large $N_c$) provided one assumes
$\Gamma_Y(\bmx_i,\bmx_j)$ to be a {\em separable} function of $Y$ and the
transverse coordinates, plus an arbitrary function of $Y$. This property is certainly not satisfied in general, but it is locally satisfied under some approximations --- e.g., in the window for extended geometric scaling, where
$\Gamma_Y(r)\simeq (r^2 Q_s^2(Y))^{\gamma_s}$ with $\gamma_s\approx 0.63$
(modulo more slowly varying logarithmic terms)
\cite{Iancu:2002tr,Mueller:2002zm,Munier:2003vc}, and deep at saturation where
$\Gamma_Y(r)\simeq (1/2\omega)\ln^2 [r^2 Q_s^2(Y)]$ (cf.~\eqn{SLT}). This is also fulfilled in some dipole models, like the GBW model
\cite{GolecBiernat:1998js,GolecBiernat:1999qd}, where $\Gamma_Y(r)=r^2
Q_s^2(Y)/4$. Assuming this to be the case, then it is possible to
explicitly perform the integral over $y$ in \eqn{QsolS} and thus obtain,
after some algebraic manipulations,
 \beq\label{Qgaussian}
 \hspace*{-0.5cm}
 \lan \hat{Q}_{\bmx_1\bmx_2\bmx_3\bmx_4} \ran_Y &=&
 \frac{\Gamma_Y(\bmx_1,\bmx_2)+
 \Gamma_Y(\bmx_3,\bmx_4)-
 \Gamma_Y(\bmx_1,\bmx_3)-
 \Gamma_Y(\bmx_2,\bmx_4)}
 {\Gamma_Y(\bmx_1,\bmx_2)+
 \Gamma_Y(\bmx_3,\bmx_4) -
 \Gamma_Y(\bmx_1,\bmx_4)-
 \Gamma_Y(\bmx_2,\bmx_3)} \,
 {\lan \hat{S}_{\bmx_1\bmx_2} \ran_Y
 \lan \hat{S}_{\bmx_3\bmx_4} \ran_Y}
 \nn
 &+&
 \frac{\Gamma_Y(\bmx_1,\bmx_4)+
 \Gamma_Y(\bmx_2,\bmx_3)-
 \Gamma_Y(\bmx_1,\bmx_3)-
 \Gamma_Y(\bmx_2,\bmx_4)}
 {\Gamma_Y(\bmx_1,\bmx_4)+
 \Gamma_Y(\bmx_2,\bmx_3) -
 \Gamma_Y(\bmx_1,\bmx_2)-
 \Gamma_Y(\bmx_3,\bmx_4)} \,
 {\lan \hat{S}_{\bmx_1\bmx_4} \ran_Y
 \lan \hat{S}_{\bmx_3\bmx_2} \ran_Y}, %\,\nn
 \eeq
which is indeed the same as the corresponding expression in the MV model
at large $N_c$ \cite{JalilianMarian:2004da,Dominguez:2011wm}. Notice that
the quadrupole at $Y_0$ does not appear anymore since in writing
\eqn{Qgaussian} we have assumed that the latter is already valid at
$Y_0$,  as correct for initial conditions of the MV type.

Returning to the general formula \eqref{QsolS}, let us notice that there
are special quadrupole configurations, like those shown in
Fig.~\ref{Fig:configs}, which have enough symmetry for the integral over
$y$ to be exactly doable, without additional assumptions. For instance,
consider the case where all transverse separations between a quark and an
antiquark are the same, that is $r_{12} = r_{34} = r_{14}= r_{32}\equiv
r$. (This includes the `line', Fig.~\ref{Fig:configs}.a, and also the
first `square', Fig.~\ref{Fig:configs}.b.) Then the integrand in
\eqn{QsolS} is a total derivative, so one easily finds
\beq\label{equaldipoles}
\lan \hat{Q}\ran_Y = \lan \hat{S}(r)\ran_Y^2 \left[\frac{\lan
\hat{Q}\ran_{Y_0}}{\lan \hat{S}(r)\ran_{Y_0}^2} + \ln \frac{\lan
\hat{S}(r)\ran_Y^2} {\lan \hat{S}(r_{13})\ran_Y \lan
\hat{S}(r_{24})\ran_Y} - \ln \frac{\lan \hat{S}(r)\ran_{Y_0}^2} {\lan
\hat{S}(r_{13})\ran_{Y_0} \lan \hat{S}(r_{24})\ran_{Y_0}} \right].
\eeq
The arguments of $\lan \hat{Q}\ran_Y$, which here is a function of $r$,
$r_{13}$, and $r_{24}$, are kept implicit.

The `line' configuration in Fig.~\ref{Fig:configs}.a corresponds to
$r_{13}=r_{24}=0$, and then \eqn{equaldipoles} reduces to \eqn{QL}, as
expected. As for the `square' configuration in Fig.~\ref{Fig:configs}.b,
one has $r_{13} = r_{24} = \sqrt{2}r$, so \eqn{equaldipoles} gives
\beq\label{QSq1}
\hspace*{-0.5cm} \lan \hat{Q}(r)\ran_Y = \lan \hat{S}(r)\ran_Y^2
\left[\frac{\lan \hat{Q}(r)\ran_{Y_0}}{\lan \hat{S}(r)\ran_{Y_0}^2} + \ln
\frac{\lan \hat{S}(r)\ran_Y^2} {\lan \hat{S}(\sqrt{2}r)\ran_Y^2} - \ln
\frac{\lan \hat{S}(r)\ran_{Y_0}^2} {\lan \hat{S}(\sqrt{2}r)\ran_{Y_0}^2}
\right]\,\,\mbox{(`square' in Fig.~\ref{Fig:configs}.b)}\,.
\eeq
Notice that the above reduce to
\beq\label{QSq1simple}
\lan \hat{Q}(r)\ran_Y = \lan \hat{S}(r)\ran_Y^2 \left[1 + \ln \frac{\lan
\hat{S}(r)\ran_Y^2} {\lan \hat{S}(\sqrt{2}r)\ran_Y^2} \right],
\eeq
provided \eqn{QSq1simple} is also valid at $Y_0$, as it is indeed the
case in the MV model at large-$N_c$. More generally, \eqn{QSq1simple} is
valid at $Y$ even if it was {\em not} valid at $Y_0$ provided the
scattering of a dipole with the given size $r$ was weak at $Y_0$. (But of
course, the scattering can be strong at $Y > Y_0$ for the same size $r$.)
Indeed, assuming that $\lan \hat{T}(r) \ran_{Y_0}\ll 1$, and therefore
also $\lan \hat{T}(\sqrt{2}r) \ran_{Y_0}\ll 1$, it is straightforward to
show that the pieces linear in $\lan \hat{T} \ran_{Y_0}$ coming from the
first and the last term of \eqn{QSq1} mutually cancel and then
\eqn{QSq1simple} is again obtained.

In Ref.~\cite{Dumitru:2011vk}, \eqn{QSq1simple} has been compared to the
numerical solution to JIMWLK equation and the agreement appears to be
remarkably good --- actually, even better than for the `line'
configuration. Clearly, expressions like \eqref{QSq1simple} or
\eqref{QL}, which involve a single transverse scale $r$, exhibit
geometric scaling in the same regime where the dipole does so. This
feature too is in agreement with the numerical results in
Ref.~\cite{Dumitru:2011vk}

A second class of configurations for which the integrand in \eqn{QsolS}
is a total derivative are those which are constrained by $r_{13}=r_{14}$
and $r_{23}=r_{24}$, whereas the two other separations $r_{12}$ and
$r_{34}$ are left arbitrary. The two last configurations in
Fig.~\ref{Fig:configs} --- the  second `square',
Fig.~\ref{Fig:configs}.c, and the `double triangle',
Fig.~\ref{Fig:configs}.d --- belong to this category. For any
configuration in this class, \eqn{QsolS} yields a very simple,
factorized, expression
 \beq\label{cancel}
 \lan \hat{Q} \ran_Y =\frac{\lan \hat{Q} \ran_{Y_0}}
 {\lan \hat{S}(r_{12})\ran_{Y_0}
 \lan \hat{S}(r_{34})\ran_{Y_0}}\,
 \lan \hat{S}(r_{12})\ran_Y
 \lan \hat{S}(r_{34})\ran_Y,
 \eeq
which involves the unconstrained separations  $r_{12}$ and $r_{34}$, but
it is independent of the other ones (those which are constrained). As
before, \eqn{cancel} reduces to an even simpler expression
  \beq\label{triangle}
 \lan \hat{Q} \ran_Y =
 \lan \hat{S}(r_{12})\ran_Y
 \lan \hat{S}(r_{34})\ran_Y,
 \eeq
provided this last formula holds already at $Y_0$ (a property which, once
again, is satisfied within the MV model at large $N_c$). In particular,
for the second `square' configuration in Fig.~\ref{Fig:configs}.c, one
has $r_{13} = r_{14} = r_{23} = r_{24} \equiv r$ and $r_{12} = r_{34} =
\sqrt{2}\, r$, and \eqn{triangle} gives
 \beq\label{othersquare}
 \lan \hat{Q}(r)\ran_Y
 %\frac{\lan \hat{Q}(r)\ran_{Y_0}}{\lan \hat{S}(\sqrt{2}r)\ran_{Y_0}^2}
 %\lan \hat{S}(\sqrt{2}\,r)\ran_Y^2
 \,=\,\lan \hat{S}(\sqrt{2}\,r)\ran_Y^2
 \qquad\mbox{(`square' in Fig.~\ref{Fig:configs}.c)}\,.
 \eeq

But the most interesting configuration in our opinion is the last one in
Fig.~\ref{Fig:configs}: the `double triangle'. What is remarkable about
this configuration is the fact that the separation between the $q\bar q$
pair $(\bmx_1,\bmx_2)$ and the second $q\bar q$ pair $(\bmx_3,\bmx_4)$
can be made arbitrary, yet the corresponding $S$--matrix in \eqn{cancel}
is independent of this separation. In particular, so long as $r_{12}$ and
$r_{34}$ are much smaller than $1/Q_s$, the scattering of the quadrupole
as a whole remains weak {\em independently} of the distance $\sim r_{14}$
separating the two pairs --- including in the case where $r_{14}\gg
1/Q_s(Y)$. This may look counterintuitive since normally one expects the
scattering to be strong whenever the separation between color charges is
of order $1/Q_s$ or larger. However, we believe that this result can be
physically understood as follows: when the two quark--antiquark pairs
$(\bmx_1,\bmx_2)$ and respectively $(\bmx_3,\bmx_4)$ are individually
small, whereas the distance $\sim r_{14}$ between them is of order
$1/Q_s$ or larger, the color exchanges between the two pairs are strongly
suppressed by saturation; then, the color charges compensates {\em
locally}, within each of the small $q\bar q$ pairs, which therefore
behave like two individual dipoles. In view of this argument, the
factorized structure of \eqn{cancel} looks quite natural. It would be
very interesting to verify this formula via numerical simulations of the
JIMWLK evolution, along the lines in Ref.~\cite{Dumitru:2011vk}.

\section*{Acknowledgments}
We would like to thank Adrian Dumitru, Giovanni Chirilli, and Al Mueller
for useful discussions. Figures were made with Jaxodraw
\cite{Binosi:2003yf,Binosi:2008ig}.

\appendix

\section{The evolution equation for the quadrupole}
\label{sect:derivation}

In this section, we shall present a particularly streamlined derivation
of the evolution equation for the quadrupole, using the `dipole' version
of the JIMWLK Hamiltonian in \eqn{H}. (An alternative derivation can be
found in Ref.~\cite{Dominguez:2011gc}.)

For pedagogy, we shall first derive the evolution equation for  a color
dipole, that is, the 2--point function of the Wilson lines shown in
\eqn{Sdipole}. The action of the functional derivative on these Wilson
lines is computed according to \eqn{donV}, thus yielding
 \beq\label{ddonVV}
 \frac{\delta}{\delta \alpha^a_{\bmu}}\,
 \frac{\delta}{\delta \alpha^b_{\bmv}}\,
 V^{\dagger}_{\bmx_1} V_{\bmx_2}
 \rightarrow
 g^2 \delta_{\bmx_1\bmu} \delta_{\bmx_2\bmv}\,
 t^a V_{\bmx_1}^{\dagger} V_{\bmx_2} t^b
 + g^2 \delta_{\bmx_1\bmv} \delta_{\bmx_2\bmu}\,
 t^b V_{\bmx_1}^{\dagger} V_{\bmx_2} t^a,
 \eeq
where in writing the r.h.s. we have dropped some terms proportional to
$\delta_{\bmu\bmv}$ since the kernel $\mcal{M}_{\bmu\bmv\bmz}$ vanishes
when $\bmu=\bmv$.

% \beq\label{adjtofun}
%\big(\wt{V}^{\dagger}\big)^{ac} t^a =
 %\wt{V}^{cb} t^b = V^{\dagger} t^c\, V,
 %\eeq

Consider the first term in the parenthesis of \eqn{H}, that is, the
identity matrix. For it, both terms arising from \eqn{ddonVV} contribute
equally and together yield the following contribution to $H
\hat{S}_{\bmx_1\bmx_2}$ :
 \beq\label{first}
 -\frac{N_c^2-1}{2 N_c^2}\,
 \atpi\,
 \int\limits_{\bmz}
 \mcal{M}_{\bmx_1\bmx_2\bmz}\, \hat{S}_{\bmx_1\bmx_2},
 \eeq
where we have also used $t^a t^a = (N_c^2-1)/2 N_c$. Next consider the
contribution coming from the second term in the parenthesis of \eqn{H}.
After expressing adjoint Wilson lines in terms of fundamental ones
according to
 \beq\label{adjtofun}
 \big(\wt{V}^{\dagger}\big)^{ac} t^a =
 \wt{V}^{cb} t^b = V^{\dagger} t^c\, V,
 \eeq
it becomes straightforward to show that this second term yields the same
contribution as the first one, that is, \eqn{first}. The third term in
the Hamiltonian leads to contributions like
 \beq
 \big(\wt{V}_{\bmx_1}^{\dagger}\big)^{ac}\,
 \wt{V}_{\bmz}^{cb}\,
 \rmtr \big(t^a\, V_{\bmx_1}^{\dagger} V_{\bmx_2} t^b  \big) =
 \rmtr\big(V_{\bmx_1}^{\dagger}\, t^c\, V_{\bmx_2}
  V_{\bmz}^{\dagger} \,t^c\, V_{\bmz} \big).
 \eeq
By using the Fierz identity
 \beq\label{Fierz1}
 \rmtr \big(t^a A \,t^a B \big)
 =\frac{1}{2}\,\rmtr A \, \rmtr B
 -\frac{1}{2 N_c}\, \rmtr(AB),
 %\nn \rmtr \big(t^a A \big)
 %\rmtr \big(t^a B \big)
 %&=&\frac{1}{2}\,\rmtr(A B)
 %-\frac{1}{2 N_c}\, \rmtr(A) \rmtr(B),
 \eeq
we find that the corresponding contribution to $H
\hat{S}_{\bmx_1\bmx_2}$, which also is equal to the one coming from the
fourth term in \eqn{H}, is
 \beq\label{third}
 \frac{1}{2}\,\atpi\,
 \int\limits_{\bm{z}}
 \mcal{M}_{\bmx_1\bmx_2\bmz}
 \Big(\hat{S}_{\bmx_1\bmz} \hat{S}_{\bmz\bmx_2}
 -\frac{1}{N_c^2}\, \hat{S}_{\bmx_1\bmx_2}\Big).
 \eeq
Putting all contributions together and averaging over the color field of
the target, we arrive at the dipole equation shown in \eqn{BK}.

Now let us turn our attention to the object of interest, the quadrupole
operator, as defined in \eqn{Squadrupole}. The two functional derivatives
can now act on 6 pairs of Wilson lines while the remaining two Wilson
lines are simply `spectators'. Just for the sake of illustration we shall
give here some intermediate steps for the terms that arise when the
`active' pair is the one composed of the two quarks at $\bmx_1$ and
$\bmx_3$. Acting on the respective Wilson lines with the first term of
the Hamiltonian, we find
 \beq
 \frac{g^2}{8 \pi^3 N_c}
 \int_{\bmz} \mcal{M}_{\bmx_1\bmx_3\bmz}
 \rmtr(t^a {V}^{\dagger}_{\bmx_1} {V}_{\bmx_2}
 t^a {V}^{\dagger}_{\bmx_3} {V}_{\bmx_4}),\nonumber
 \eeq
which after simple manipulations using the Fierz identity in \eqn{Fierz1}
leads to
 \beq\label{Qfirst}
 \frac{\abar}{4\pi} \int_{\bmz}
 \mcal{M}_{\bmx_1\bmx_3\bmz}
 \Big(\hat{S}_{\bmx_1\bmx_2}\hat{S}_{\bmx_3\bmx_4}
 -\frac{1}{N_c^2}\, \hat{Q}_{\bmx_1\bmx_2\bmx_3\bmx_4}\Big).
 \eeq
Acting with the second term of the Hamiltonian we obtain
 \beq
 \frac{g^2}{16 \pi^3 N_c}
 \int_{\bmz} \mcal{M}_{\bmx_1\bmx_3\bmz}
 \left[
 \big(\wt{V}_{\bmx_1}^{\dagger}\big)^{ac} \wt{V}_{\bmx_3}^{cb}\,
 \rmtr(t^a {V}^{\dagger}_{\bmx_1} {V}_{\bmx_2}
 t^b {V}^{\dagger}_{\bmx_3} {V}_{\bmx_4})
 +
 \big(\wt{V}_{\bmx_3}^{\dagger}\big)^{ac} \wt{V}_{\bmx_1}^{cb}\,
 \rmtr(t^b {V}^{\dagger}_{\bmx_1} {V}_{\bmx_2}
 t^a {V}^{\dagger}_{\bmx_3} {V}_{\bmx_4})\right],\nonumber
 \eeq
and by also making use of \eqn{adjtofun} we are lead to
 \beq\label{Qsecond}
 \frac{\abar}{4\pi} \int_{\bmz}
 \mcal{M}_{\bmx_1\bmx_3\bmz}
 \Big(\hat{S}_{\bmx_3\bmx_2}\hat{S}_{\bmx_1\bmx_4}
 -\frac{1}{N_c^2}\, \hat{Q}_{\bmx_1\bmx_2\bmx_3\bmx_4}\Big).
 \eeq
The action of the third term of the Hamiltonian gives
 \beq
 -\frac{g^2}{16 \pi^3 N_c}
 \int_{\bmz} \mcal{M}_{\bmx_1\bmx_3\bmz}
 \left[
 \big(\wt{V}_{\bmx_1}^{\dagger}\big)^{ac} \wt{V}_{\bmz}^{cb}\,
 \rmtr(t^a {V}^{\dagger}_{\bmx_1} {V}_{\bmx_2}
 t^b {V}^{\dagger}_{\bmx_3} {V}_{\bmx_4})
 +
 \big(\wt{V}_{\bmx_3}^{\dagger}\big)^{ac} \wt{V}_{\bmz}^{cb}\,
 \rmtr(t^b {V}^{\dagger}_{\bmx_1} {V}_{\bmx_2}
 t^a {V}^{\dagger}_{\bmx_3} {V}_{\bmx_4})\right],\nonumber
 \eeq
leading to
 \beq\label{Qthird}
 -\frac{\abar}{8\pi} \int_{\bmz}
 \mcal{M}_{\bmx_1\bmx_3\bmz}
 \Big(\hat{S}_{\bmz\bmx_2}\hat{Q}_{\bmx_1\bmz\bmx_3\bmx_4}
 + \hat{S}_{\bmz\bmx_4}\hat{Q}_{\bmx_1\bmx_2\bmx_3\bmz}
 -\frac{2}{N_c^2}\, \hat{Q}_{\bmx_1\bmx_2\bmx_3\bmx_4}\Big).
 \eeq
Finally the action of the fourth term in the Hamiltonian gives
 \beq
 -\frac{g^2}{16 \pi^3 N_c}
 \int_{\bmz} \mcal{M}_{\bmx_1\bmx_3\bmz}
 \left[
 \big(\wt{V}_{\bmz}^{\dagger}\big)^{ac} \wt{V}_{\bmx_3}^{cb}\,
 \rmtr(t^a {V}^{\dagger}_{\bmx_1} {V}_{\bmx_2}
 t^b {V}^{\dagger}_{\bmx_3} {V}_{\bmx_4})
 +
 \big(\wt{V}_{\bmz}^{\dagger}\big)^{ac} \wt{V}_{\bmx_1}^{cb}\,
 \rmtr(t^b {V}^{\dagger}_{\bmx_1} {V}_{\bmx_2}
 t^a {V}^{\dagger}_{\bmx_3} {V}_{\bmx_4})\right],\nonumber
 \eeq
yielding the same result as shown in \eqn{Qthird}. Putting together all
the contributions, we see that the terms explicitly suppressed by
$1/N_c^2$ cancel each other, as anticipated. Thus, the action of the
Hamiltonian on the pair made with the two quarks at $\bmx_1$ and $\bmx_3$
gives
 \beq\label{Q13}
 -\frac{\abar}{4\pi} \int_{\bmz}
 \mcal{M}_{\bmx_1\bmx_3\bmz}
 \big(\hat{S}_{\bmz\bmx_2}\hat{Q}_{\bmx_1\bmz\bmx_3\bmx_4}
 + \hat{S}_{\bmz\bmx_4}\hat{Q}_{\bmx_1\bmx_2\bmx_3\bmz}
 -\hat{S}_{\bmx_1\bmx_2}\hat{S}_{\bmx_3\bmx_4}
 -\hat{S}_{\bmx_3\bmx_2}\hat{S}_{\bmx_1\bmx_4} \big).
 \eeq
Similarly, when acting on the quark and the antiquark at $\bmx_1$ and
$\bmx_2$ respectively, we obtain
 \beq\label{Q12}
 \frac{\abar}{4\pi} \int_{\bmz}
 \mcal{M}_{\bmx_1\bmx_2\bmz}
 \big(\hat{S}_{\bmx_1\bmz}\hat{Q}_{\bmz\bmx_2\bmx_3\bmx_4}
 + \hat{S}_{\bmz\bmx_2}\hat{Q}_{\bmx_1\bmz\bmx_3\bmx_4}
 -\hat{Q}_{\bmx_1\bmx_2\bmx_3\bmx_4}
 -\hat{S}_{\bmx_1\bmx_2}\hat{S}_{\bmx_3\bmx_4} \big).
 \eeq
The action of the Hamiltonian on the other possible pairs of Wilson lines
is obtained via appropriate permutations in either \eqn{Q13} or
\eqn{Q12}. Putting all terms together and averaging over the target
field, we arrive at the evolution equation shown in \eqn{Qevol} in the
main text.

\section{The sextupole}
\label{sect:sextupole}

In order to demonstrate the generality of our procedure, we shall
succinctly present here the corresponding analysis for the next
non--trivial high--point correlator, which is the sextupole defined in
\eqn{Ssextupole}. The respective evolution equation has not been yet
presented in the literature, thus we have derived it on this occasion. We
shall not give here the details of the derivation, since this is merely a
lengthy but straightforward repetition of the manipulations shown in the
previous Appendix. This is the final result:
 \beq\label{S6evol}
 \hspace*{-.4cm}
 \frac{\del \lan\hat{S}_{\bmx_1\bmx_2\bmx_3\bmx_4\bmx_5\bmx_6}^{(6)}
 \ran_Y}{\del Y} &=& \frac{\abar}{4\pi} \int_{\bmz}
 \Big[(\mcal{M}_{\bmx_1\bmx_2\bmz}+ \mcal{M}_{\bmx_1\bmx_6\bmz}
 -\mcal{M}_{\bmx_2\bmx_6\bmz})
 \lan \hat{S}_{\bmx_1 \bmz} \hat{S}_{\bmz\bmx_2\bmx_3\bmx_4\bmx_5\bmx_6}^{(6)} \ran_Y
 \nn
 &+&(\mcal{M}_{\bmx_1\bmx_2\bmz}+ \mcal{M}_{\bmx_2\bmx_3\bmz}
 -\mcal{M}_{\bmx_1\bmx_3\bmz})
 \lan \hat{S}_{\bmz \bmx_2} \hat{S}_{\bmx_1\bmz\bmx_3\bmx_4\bmx_5\bmx_6}^{(6)} \ran_Y
 \nn
 &+&(\mcal{M}_{\bmx_2\bmx_3\bmz}+ \mcal{M}_{\bmx_3\bmx_4\bmz}
 -\mcal{M}_{\bmx_2\bmx_4\bmz})
 \lan \hat{S}_{\bmx_3 \bmz} \hat{S}_{\bmx_1\bmx_2\bmz\bmx_4\bmx_5\bmx_6}^{(6)} \ran_Y
 \nn
 &+&(\mcal{M}_{\bmx_3\bmx_4\bmz}+ \mcal{M}_{\bmx_4\bmx_5\bmz}
 -\mcal{M}_{\bmx_3\bmx_5\bmz})
 \lan \hat{S}_{\bmz \bmx_4} \hat{S}_{\bmx_1\bmx_2\bmx_3\bmz\bmx_5\bmx_6}^{(6)} \ran_Y
 \nn
 &+&(\mcal{M}_{\bmx_4\bmx_5\bmz}+ \mcal{M}_{\bmx_5\bmx_6\bmz}
 -\mcal{M}_{\bmx_4\bmx_6\bmz})
 \lan \hat{S}_{\bmx_5 \bmz} \hat{S}_{\bmx_1\bmx_2\bmx_3\bmx_4\bmz\bmx_6}^{(6)} \ran_Y
 \nn
 &+&(\mcal{M}_{\bmx_1\bmx_6\bmz}+ \mcal{M}_{\bmx_5\bmx_6\bmz}
 -\mcal{M}_{\bmx_1\bmx_5\bmz})
 \lan \hat{S}_{\bmz \bmx_6} \hat{S}_{\bmx_1\bmx_2\bmx_3\bmx_4\bmx_5\bmz}^{(6)} \ran_Y
 \nn
 &+&(\mcal{M}_{\bmx_1\bmx_4\bmz}+ \mcal{M}_{\bmx_3\bmx_6\bmz}
 -\mcal{M}_{\bmx_1\bmx_3\bmz}-\mcal{M}_{\bmx_4\bmx_6\bmz})
 \lan \hat{Q}_{\bmx_1 \bmx_2 \bmx_3 \bmz} \hat{Q}_{\bmz \bmx_4\bmx_5\bmx_6} \ran_Y
 \nn
 &+&(\mcal{M}_{\bmx_2\bmx_5\bmz}+ \mcal{M}_{\bmx_3\bmx_6\bmz}
 -\mcal{M}_{\bmx_3\bmx_5\bmz}-\mcal{M}_{\bmx_2\bmx_6\bmz})
 \lan \hat{Q}_{\bmx_3 \bmx_4 \bmx_5 \bmz} \hat{Q}_{\bmz \bmx_6\bmx_1\bmx_2}  \ran_Y
 \nn
 &+&(\mcal{M}_{\bmx_1\bmx_4\bmz}+ \mcal{M}_{\bmx_2\bmx_5\bmz}
 -\mcal{M}_{\bmx_1\bmx_5\bmz}-\mcal{M}_{\bmx_2\bmx_4\bmz})
 \lan\hat{Q}_{\bmx_5\bmx_6\bmx_1\bmz}
 \hat{Q}_{\bmz \bmx_2 \bmx_3 \bmx_4} \ran_Y
 \nn
 &-&(\mcal{M}_{\bmx_1\bmx_2\bmz} + \mcal{M}_{\bmx_2\bmx_3\bmz}
 +\mcal{M}_{\bmx_3\bmx_4\bmz} + \mcal{M}_{\bmx_4\bmx_5\bmz}
 +\mcal{M}_{\bmx_5\bmx_6\bmz} + \mcal{M}_{\bmx_6\bmx_1\bmz})
 \lan\hat{S}_{\bmx_1\bmx_2\bmx_3\bmx_4\bmx_5\bmx_6}^{(6)} \ran_Y
 \nn
 &-&(\mcal{M}_{\bmx_1\bmx_2\bmz} + \mcal{M}_{\bmx_3\bmx_6\bmz}
 -\mcal{M}_{\bmx_1\bmx_3\bmz} - \mcal{M}_{\bmx_2\bmx_6\bmz})
 \lan
 \hat{S}_{\bmx_1\bmx_2}
 \hat{Q}_{\bmx_3\bmx_4\bmx_5\bmx_6}
 \ran_Y
 \nn
 &-&(\mcal{M}_{\bmx_1\bmx_4\bmz} + \mcal{M}_{\bmx_2\bmx_3\bmz}
 -\mcal{M}_{\bmx_1\bmx_3\bmz} - \mcal{M}_{\bmx_2\bmx_4\bmz})
 \lan
 \hat{S}_{\bmx_3\bmx_2}
 \hat{Q}_{\bmx_1\bmx_4\bmx_5\bmx_6}
 \ran_Y
 \nn
 &-&(\mcal{M}_{\bmx_3\bmx_4\bmz} + \mcal{M}_{\bmx_2\bmx_5\bmz}
 -\mcal{M}_{\bmx_2\bmx_4\bmz} - \mcal{M}_{\bmx_3\bmx_5\bmz})
 \lan
 \hat{S}_{\bmx_3\bmx_4}
 \hat{Q}_{\bmx_1\bmx_2\bmx_5\bmx_6}
 \ran_Y
 \nn
 &-&(\mcal{M}_{\bmx_3\bmx_6\bmz} + \mcal{M}_{\bmx_4\bmx_5\bmz}
 -\mcal{M}_{\bmx_3\bmx_5\bmz} - \mcal{M}_{\bmx_4\bmx_6\bmz})
 \lan
 \hat{S}_{\bmx_5\bmx_4}
 \hat{Q}_{\bmx_1\bmx_2\bmx_3\bmx_6}
 \ran_Y
 \nn
 &-&(\mcal{M}_{\bmx_1\bmx_4\bmz} + \mcal{M}_{\bmx_5\bmx_6\bmz}
 -\mcal{M}_{\bmx_1\bmx_5\bmz} - \mcal{M}_{\bmx_4\bmx_6\bmz})
 \lan
 \hat{S}_{\bmx_5\bmx_6}
 \hat{Q}_{\bmx_1\bmx_2\bmx_3\bmx_4}
 \ran_Y
 \nn
 &-&(\mcal{M}_{\bmx_1\bmx_6\bmz} + \mcal{M}_{\bmx_2\bmx_5\bmz}
 -\mcal{M}_{\bmx_1\bmx_5\bmz} - \mcal{M}_{\bmx_2\bmx_6\bmz})
 \lan
 \hat{S}_{\bmx_1\bmx_6}
 \hat{Q}_{\bmx_3\bmx_4\bmx_5\bmx_2}
 \ran_Y\Big].
 \eeq
We notice that all the subleading terms of relative order $1/N_c^2$ have
canceled in the final equation, as expected for a single trace operator.
As a cross check, one can see that when choosing, for example, $\bmx_5 =
\bmx_6$, the above equation reduces to the quadrupole equation
\eqref{Qevol}.

The subsequent manipulations follow the general procedure outlined in
Sect.~\ref{sect:quadru}. First, one separates the `virtual' terms in the
r.h.s. of \eqn{S6evol}, that is, the terms in which the integration
variable $\bmz$ appears only in the dipole kernel, but not in the
correlation functions; these are the terms proportional to $\lan
\hat{S}^{(6)} \ran_Y$ and to  $\lan \hat{S} \hat{Q} \ran_Y \simeq \lan
\hat{S}\ran_Y  \lan \hat{Q} \ran_Y$. Then in evaluating the integrals
over  $\bmz$, one keeps only the dominant logarithmic contributions and
use \eqn{BKstrong} to express the logarithms in term of the dipole
derivative. We thus arrive at an ordinary first order inhomogeneous
differential equation whose solution is
 \beq\label{S6solQS}
 \lan\hat{S}_{\bmx_1\bmx_2\bmx_3\bmx_4\bmx_5\bmx_6}^{(6)}\ran_Y &=&
 \sqrt{\lan \hat{S}_{\bmx_1\bmx_2} \ran_Y
 \lan \hat{S}_{\bmx_3\bmx_2} \ran_Y
 \lan \hat{S}_{\bmx_3\bmx_4} \ran_Y
 \lan \hat{S}_{\bmx_5\bmx_4} \ran_Y
 \lan \hat{S}_{\bmx_5\bmx_6} \ran_Y
 \lan \hat{S}_{\bmx_1\bmx_6} \ran_Y}
 \nn
 &{}&\ \times\left\{
 \frac{\lan\hat{S}_{\bmx_1\bmx_2\bmx_3\bmx_4\bmx_5\bmx_6}^{(6)} \ran_{Y_0}}
 {\sqrt{\lan \hat{S}_{\bmx_1\bmx_2} \ran_{Y_0}
 \lan \hat{S}_{\bmx_3\bmx_2} \ran_{Y_0}
 \lan \hat{S}_{\bmx_3\bmx_4} \ran_{Y_0}
 \lan \hat{S}_{\bmx_5\bmx_4} \ran_{Y_0}
 \lan \hat{S}_{\bmx_5\bmx_6} \ran_{Y_0}
 \lan \hat{S}_{\bmx_1\bmx_6} \ran_{Y_0}}}\right.
 \nn
 &{}& \hspace*{0.3cm} +\frac{1}{2}\,\int_{Y_0}^{Y} \dif y\,
 \frac{1}
 {\sqrt{\lan \hat{S}_{\bmx_1\bmx_2} \ran_y
 \lan \hat{S}_{\bmx_3\bmx_2} \ran_y
 \lan \hat{S}_{\bmx_3\bmx_4} \ran_y
 \lan \hat{S}_{\bmx_5\bmx_4} \ran_y
 \lan \hat{S}_{\bmx_5\bmx_6} \ran_y
 \lan \hat{S}_{\bmx_1\bmx_6} \ran_y}}
 \nn
 &{}& \hspace*{0.6cm}\times \Bigg[ \lan \hat{S}_{\bmx_1\bmx_2} \ran_y
 \lan \hat{Q}_{\bmx_3\bmx_4\bmx_5\bmx_6} \ran_y
 \,\frac{\del}{\del y}\ln
 \frac{\lan \hat{S}_{\bmx_1\bmx_2} \ran_y
 \lan \hat{S}_{\bmx_3\bmx_6} \ran_y}
 {\lan \hat{S}_{\bmx_1\bmx_3} \ran_y
 \lan \hat{S}_{\bmx_2\bmx_6} \ran_y}
 \nn
 &{}& \hspace*{0.8cm} + \lan \hat{S}_{\bmx_3\bmx_2} \ran_y
 \lan \hat{Q}_{\bmx_1\bmx_4\bmx_5\bmx_6} \ran_y
 \,\frac{\del}{\del y}\ln
 \frac{\lan \hat{S}_{\bmx_3\bmx_2} \ran_y
 \lan \hat{S}_{\bmx_1\bmx_4} \ran_y}
 {\lan \hat{S}_{\bmx_1\bmx_3} \ran_y
 \lan \hat{S}_{\bmx_2\bmx_4} \ran_y}
 \nn
 &{}& \hspace*{0.8cm} +
 \lan \hat{S}_{\bmx_3\bmx_4} \ran_y
 \lan \hat{Q}_{\bmx_1\bmx_2\bmx_5\bmx_6} \ran_y
 \,\frac{\del}{\del y}\ln
 \frac{\lan \hat{S}_{\bmx_3\bmx_4} \ran_y
 \lan \hat{S}_{\bmx_5\bmx_2} \ran_y}
 {\lan \hat{S}_{\bmx_2\bmx_4} \ran_y
 \lan \hat{S}_{\bmx_3\bmx_5} \ran_y}
 \nn
 &{}& \hspace*{0.8cm} +
 \lan \hat{S}_{\bmx_5\bmx_4} \ran_y
 \lan \hat{Q}_{\bmx_1\bmx_2\bmx_3\bmx_6} \ran_y
 \,\frac{\del}{\del y}\ln
 \frac{\lan \hat{S}_{\bmx_5\bmx_4} \ran_y
 \lan \hat{S}_{\bmx_3\bmx_6} \ran_y}
 {\lan \hat{S}_{\bmx_3\bmx_5} \ran_y
 \lan \hat{S}_{\bmx_4\bmx_6} \ran_y}
 \nn
 &{}& \hspace*{0.8cm} +
 \lan \hat{S}_{\bmx_5\bmx_6} \ran_y
 \lan \hat{Q}_{\bmx_1\bmx_2\bmx_3\bmx_4} \ran_y
 \,\frac{\del}{\del y}\ln
 \frac{\lan \hat{S}_{\bmx_5\bmx_6} \ran_y
 \lan \hat{S}_{\bmx_1\bmx_4} \ran_y}
 {\lan \hat{S}_{\bmx_1\bmx_5} \ran_y
 \lan \hat{S}_{\bmx_4\bmx_6} \ran_y}
 \nn &{}& \hspace*{0.8cm}+\!\left.
 \lan \hat{S}_{\bmx_1\bmx_6} \ran_y
 \lan \hat{Q}_{\bmx_3\bmx_4\bmx_5\bmx_2} \ran_y
 \,\frac{\del}{\del y}\ln
 \frac{\lan \hat{S}_{\bmx_1\bmx_6} \ran_y
 \lan \hat{S}_{\bmx_5\bmx_2} \ran_y}
 {\lan \hat{S}_{\bmx_1\bmx_5} \ran_y
 \lan \hat{S}_{\bmx_2\bmx_6} \ran_y}\Bigg]\right\}.
 \eeq
Again, one can check that choosing, for example, $\bmx_5 = \bmx_6$, the
above solution reduces to the quadrupole one given in \eqn{QsolS}.
Expanding \eqn{S6solQS} in the weak scattering region, we find
 \beq\label{S6BFKL}
 \hspace*{-0.5cm}
 1 - \lan\hat{S}_{\bmx_1\bmx_2\bmx_3\bmx_4\bmx_5\bmx_6}^{(6)} \ran_Y
 &\simeq&
 \lan
 \hat{T}_{\bmx_1\bmx_2}
 -\hat{T}_{\bmx_1\bmx_3}
 +\hat{T}_{\bmx_1\bmx_4}
 -\hat{T}_{\bmx_1\bmx_5}
 +\hat{T}_{\bmx_1\bmx_6}
 +\hat{T}_{\bmx_2\bmx_3}
 -\hat{T}_{\bmx_2\bmx_4}
 +\hat{T}_{\bmx_2\bmx_5}
 \nn
 &{}& -\hat{T}_{\bmx_2\bmx_6}
 +\hat{T}_{\bmx_3\bmx_4}
 -\hat{T}_{\bmx_3\bmx_5}
 +\hat{T}_{\bmx_3\bmx_6}
 +\hat{T}_{\bmx_4\bmx_5}
 -\hat{T}_{\bmx_4\bmx_6}
 +\hat{T}_{\bmx_5\bmx_6}
 \ran_Y,
 \eeq
where we have also used \eqn{QBFKL} for $\lan\hat{Q}\ran_Y$ and we have
assumed that \eqn{S6BFKL} is already valid at $Y_0$. It is
straightforward to confirm that the same result is obtained when we
expand \eqn{Ssextupole} to order $g^2 \alpha^2$.

As a particular example, we shall consider the `line' configuration,
already studied in the case of the quadrupole. This is obtained by
putting all the quarks at the same point, $\bmx_1 = \bmx_3 = \bmx_5$, and
similarly for the anti--quarks, $\bmx_2=\bmx_4=\bmx_6$, with the two
points separated by a distance $r$. (One can trivially visualize this
configuration by looking at Fig.~\ref{Fig:configs}.a and adding there
$\bmx_5$ and $\bmx_6$ at the left and right ends, respectively.) Then by
also using \eqn{QlineS} we see that the $y$--integration in \eqn{S6solQS}
can be exactly performed and gives
 \beq\label{S6line}
 \lan \hat{S}^{(6)}\ran_Y =
 \lan \hat{S}\ran_Y^3
 \left[
 \frac{\lan \hat{S}^{(6)}\ran_{Y_0}}{\lan \hat{S}\ran_{Y_0}^3}
 + 3 \left(
 \frac{\lan \hat{Q}\ran_{Y_0}}{\lan \hat{S}\ran_{Y_0}^2}
 -\ln \lan \hat{S}\ran_{Y_0}^2
 \right)
 \ln \frac{\lan \hat{S}\ran_Y^2}{\lan \hat{S}\ran_{Y_0}^2}
 + \frac{3}{2} \ln^2 \lan \hat{S}\ran_Y^2
 - \frac{3}{2} \ln^2 \lan \hat{S}\ran_{Y_0}^2
 \right],\ \
 \eeq
where clearly all quantities are evaluated at $r$. The above simplifies
to
 \beq\label{S6linesimple}
 \lan \hat{S}^{(6)}\ran_Y =
 \lan \hat{S}\ran_Y^3
 \left[1
 + 3 \ln\lan \hat{S}\ran_Y^2
 + \frac{3}{2} \ln^2 \lan \hat{S}\ran_Y^2
 \right],
 \eeq
if we assume that the above and \eqn{QL} are already valid at $Y_0$ or if
we assume that the scattering at $Y_0$ for the given $r$ is weak. The
sextupole in \eqn{S6linesimple} exhibits geometric scaling in the same
regime where the dipole does so.

%\bibliographystyle{utcaps}
%\bibliography{refs}

\providecommand{\href}[2]{#2}\begingroup\raggedright\endgroup

\end{document}